\documentclass{npw}
\usepackage{graphicx}
\usepackage{epsfig}
\usepackage{mathrsfs,amsmath,amssymb}

\newcommand{\lc}{\left<}
\newcommand{\rc}{\right>}
\newcommand{\lr}{\left|}
\newcommand{\rl}{\right|}
\newcommand{\lb}{\left(}
\newcommand{\rb}{\right)}
\newcommand{\ls}{\left[}
\newcommand{\rs}{\right]}
\newcommand{\Lb}{\left\{}
\newcommand{\Rb}{\right\}}
\newcommand{\ff}[1]{\frac{1}{#1}}
\newcommand{\scr}[1]{{\mathscr #1}}

\newcommand{\no}{\nonumber\\}

\newcommand{\ppar}[2]{\frac{\partial #1}{\partial #2}}
\newcommand{\mean}[3]{\left<#1\right|#2\left|#3\right>}

\begin{document}

\title{Finite-amplitude method: An extension to the covariant density functionals}

\author{
  Haozhao Liang$^{1,2,}$\email{haozhao.liang@riken.jp}~, Takashi Nakatsukasa$^{1,3}$, Zhongming Niu$^4$, and Jie Meng$^{2,5,6}$ \\
  \it $^1$RIKEN Nishina Center, Wako 351-0198, Japan\\
  \it $^2$School of Physics, Peking University, Beijing 100871, China\\
  \it $^3$Center for Computational Sciences, University of Tsukuba, Tsukuba 305-8571, Japan\\
  \it $^4$School of Physics and Material Science, Anhui University, Hefei 230601, China\\
  \it $^5$School of Physics and Nuclear Energy Engineering, Beihang University, Beijing 100191, China\\
  \it $^6$Department of Physics, University of Stellenbosch, Stellenbosch, South Africa
}
\date{}
\maketitle

\begin{abstract}
The finite-amplitude method (FAM) is one of the most promising methods for optimizing the computational performance of the random-phase approximation (RPA) calculations in deformed nuclei.
In this report, we will mainly focus on our recent progress in the self-consistent relativistic RPA established by using the FAM.
It is found that the effects of Dirac sea can be taken into account implicitly in the coordinate-space representation and the rearrangement terms due to the density-dependent couplings can be treated without extra computational costs.
\end{abstract}

\section{Introduction}

During the past decades, the covariant density functional theory (CDFT) has received wide attention due to its successful descriptions of both ground-state and excited-state properties of nuclei all over the nuclear chart.
In this report, we will mainly focus on our recent progress in the self-consistent relativistic random-phase approximation (RPA) established by using the finite-amplitude method (FAM) \cite{Liang2013}.

The density functional theory has been widely used in nuclear physics since the 1970s \cite{Bender2003}.
In particular, its covariant version \cite{Serot1986,Ring1996} takes the Lorentz invariance into account, and thus puts stringent restrictions on the number of parameters, achieving a consistent treatment of the spin degrees of freedom as well as the unification of the time-even and time-odd components.
Over the years, a large variety of nuclear phenomena have been described successfully by the CDFT \cite{Vretenar2005,Meng2006,Paar2007,Niksic2011}.

The RPA \cite{Ring1980} is one of the leading theories applicable to both low-lying excited states and giant resonances.
In the relativistic framework, the self-consistent and quantitative RPA calculations were realized after recognizing the importance of the Dirac sea \cite{Ring2001}.
From then on, great efforts along this direction have been made \cite{Paar2007}.
Recently, a fully self-consistent relativistic RPA \cite{Liang2008} has been established based on the relativistic Hartree-Fock theory \cite{Long2006}.
It is shown that not only the Gamow-Teller resonances but also the fine structure of spin-dipole resonances can be well reproduced without any readjustment of the energy functional \cite{Liang2008,Liang2012a}.
This self-consistent RPA has also been applied to evaluate the isospin symmetry-breaking corrections to the superallowed $\beta$ transitions for the unitarity test of Cabibbo-Kobayashi-Maskawa matrix \cite{Liang2009}.
Recently, the corresponding quasiparticle RPA (QRPA) \cite{Niu2013} based on the relativistic Hartree-Fock-Bogoliubov theory \cite{Long2010} has been developed.

However, the above investigations are essentially restricted within the spherical symmetry.
The conventional RPA calculations in the matrix form face a big computational challenge when the number of particle-hole (\textit{ph}) configurations $N_{ph}$ becomes huge.
So far, the self-consistent deformed (Q)RPA in the relativistic framework was only developed by Pe\~na Arteaga \textit{et al.} \cite{PenaArteaga2008}.

As a promising solution for this computational challenge, the so-called finite-amplitude method was proposed in Ref.~\cite{Nakatsukasa2007}.
In this method, the effects of residual interactions are evaluated in a numerical way by considering a finite density deviation around the ground state.
In such a way, the self-consistent RPA calculations become possible with a little extension of the static Hartree(-Fock) code.
Furthermore, by using the iterative methods for the RPA equation, the computation time is close to a linear dependence on $N_{ph}$, instead of a dependence between $N^2_{ph}$ and $N^3_{ph}$ in the diagonalization scheme \cite{Avogadro2013}.
This advantage is crucial when $N_{ph}$ becomes huge.

In the non-relativistic framework with Skyrme energy density functionals, the feasibility, accuracy, and efficiency of FAM have been demonstrated for the RPA in the three-dimensionally deformed cases in the coordinate-space representation \cite{Nakatsukasa2007,Inakura2009} and for the QRPA in the spherical \cite{Avogadro2011,Avogadro2013} and axially deformed \cite{Stoitsov2011,Hinohara2013} cases in the quasiparticle-basis representation.
Iterative algorithms for (Q)RPA solutions have also been developed recently, based on the Arnoldi process \cite{Toivanen2010,Carlsson2012} and on the conjugate gradient method \cite{Imagawa2003}.
The readers are also referred to Ref.~\cite{Nakatsukasa2012} for a recent review.

Work is now in progress for developing the self-consistent relativistic RPA by using the FAM.
In particular, special attentions should be paid to the unique features of covariant density functionals, including the effects of the Dirac sea and the rearrangement terms for the density-dependent interactions.
These rearrangement terms are usually more sophisticated than those in the Skyrme functionals, and cause heavy computations \cite{Niksic2002}.
On the other hand, the covariant density functionals hold the Lorentz invariance, which leads to the unification of their time-even and time-odd components.
This makes the modification in the ground-state code straightforward.

\section{Formalism}\label{sec:TF}

In this section, we will highlight the key formulas of the FAM in CDFT restricted to spherical nuclei, in particular, we will mark the most important points by the symbol $\bigstar$.
The detailed derivations and the meaning of notations can be found in Ref.~\cite{Liang2013}.

\subsection{Point-coupling relativistic mean-field theory}

Our starting point is the effective Lagrangian density of the point-coupling relativistic mean-field (RMF) theory \cite{Niksic2008,Zhao2010,Liang2012b}, which reads
\begin{align}
    \mathcal{L} =& \bar\psi (i\gamma^\mu\partial_\mu-M)\psi
        -\ff2\alpha_S(\bar\psi\psi)(\bar\psi\psi)-\ff2\delta_S(\partial_\nu\bar\psi\psi)(\partial^\nu\bar\psi\psi)\no
        &-\ff2\alpha_V(\bar\psi\gamma^\mu\psi)(\bar\psi\gamma_\mu\psi)
        -\ff2\alpha_{tV}(\bar\psi\vec\tau\gamma^\mu\psi)\cdot(\bar\psi\vec\tau\gamma_\mu\psi)\no
        &-e\bar\psi\gamma^\mu A_\mu\frac{(1-\tau_3)}{2}\psi-\frac{1}{4}F^{\mu\nu}F_{\mu\nu},
\end{align}
where the coupling strengths $\alpha$ are analytical functions with respect to the baryonic density, $x=\rho_b/\rho_{\rm sat}$,
\begin{subequations}\label{eq:alpha}
\begin{align}
    \alpha_S(\rho_b) &= a_S + (b_S + c_Sx)e^{-d_Sx},\\
    \alpha_V(\rho_b) &= a_V + b_V e^{-d_Vx},\\
    \alpha_{tV}(\rho_b) &= b_{tV}e^{-d_{tV}x}.
\end{align}
\end{subequations}

For the systems with spherical symmetry, the single-particle wave functions have the form of
\begin{equation}
    \psi_a(\mathbf{r})=\ff r
        \Lb \begin{array}{c}
            iG_a(r) \\ F_a(r)\hat{\sigma}\cdot\hat{\mathbf{r}}
            \end{array} \Rb
            \scr Y_a(\hat{\mathbf{r}})\chi_\ff2(q_a).
\end{equation}
Hereafter, $\lr\psi_a\rc$ indicate the single-particle wave functions in general, but $\lr\phi_a\rc$ specifically represent the eigenstates of $h_0$; index $a$ runs over all single-particle states, but indices $i,j$ ($m,n$) only run over the hole (particle) states.

$\bigstar$ Note that within this phase convention between the upper and lower components, the wave functions $G(r)$ and $F(r)$ can be simultaneously chosen as real functions for the ground-state descriptions.
In contrast, for the FAM built below, both $G(r)$ and $F(r)$ become complex functions, so one should be careful to distinguish them from their complex conjugates $G^*(r)$ and $F^*(r)$ from the very beginning.

The radial Dirac equation reads
\begin{align}\label{eq:rDirac}
    &\lb \begin{array}{cc}
        M+\Sigma_S+\Sigma_0 & -\frac{d}{dr}+\frac{\kappa_a}{r}+\Sigma_V \\
        \frac{d}{dr}+\frac{\kappa_a}{r}-\Sigma_V & -M-\Sigma_S+\Sigma_0
    \end{array} \rb
    \lb \begin{array}{c}
        G_a \\ F_a
    \end{array} \rb\nonumber\\
    &=\varepsilon_a
    \lb \begin{array}{c}
        G_a \\ F_a
    \end{array} \rb,
\end{align}
with the scalar and vector potentials
\begin{subequations}\label{eq:Sigma}
\begin{align}
    \Sigma_S(r) &= \alpha_S \rho_S + \delta_S\lb\rho''_S+\frac{2}{r}\rho'_S\rb,\\
    \Sigma_0(r) &= \alpha_V \rho_V + \alpha_{tV}\rho_{tV}\tau_3 +e\frac{1-\tau_3}{2}A_0 + \Sigma_R,\\
    \Sigma_V(r) &= \alpha_V j_V + \alpha_{tV}j_{tV}\tau_3 +e\frac{1-\tau_3}{2}A_V.
\end{align}
\end{subequations}
The rearrangement terms only contribute to the time-like component of the vector potential, which read
\begin{equation}\label{eq:SigmaR}
    \Sigma_R(r)=\ff2\Lb \ppar{\alpha_S}{\rho_b}\rho_S^2 + \ppar{\alpha_V}{\rho_b}(\rho_V^2+j_V^2)
        + \ppar{\alpha_{tV}}{\rho_b}(\rho_{tV}^2+j_{tV}^2)\Rb.
\end{equation}
The densities and currents are expressed as
\begin{subequations}\label{eq:dens}
\begin{align}
    \rho_S^{(q_a)}&=
        \ff{4\pi r^2}\sum^{q_a}v_a^2\hat j_a^2
        \ls G^*_a(r)G_a(r)-F^*_a(r)F_a(r)\rs,\\
    \rho_V^{(q_a)}&=
        \ff{4\pi r^2}\sum^{q_a}v_a^2\hat j_a^2
        \ls G^*_a(r)G_a(r)+F^*_a(r)F_a(r)\rs,\\
    j_V^{(q_a)}&=
        \ff{4\pi r^2}\sum^{q_a}v_a^2\hat j_a^2
        \ls G^*_a(r)F_a(r)-F^*_a(r)G_a(r)\rs.
\end{align}
\end{subequations}
The Coulomb fields are calculated with the Green's function method,
\begin{subequations}\label{eq:Coulomb}
\begin{align}
    A_0(r) &= e\int dr' {r'}^2 \rho_V^{(p)}(r') \ff{r_>},\\
    A_V(r) &= \frac{e}{3}\int dr' {r'}^2 j_V^{(p)}(r') \frac{r_<}{r^2_>}.
\end{align}
\end{subequations}

\subsection{Linear response and RPA}

The RPA equation is known to be equivalent to the time-dependent Hartree(-Fock) equation in the small amplitude limit \cite{Ring1980}.

The static Hartree(-Fock) equation, $[h[\rho], \rho] = 0$, determines the ground-state density $\rho=\rho_0$ and the one-body mean-field Hamiltonian $h_0 = h[\rho_0]$.

When a time-dependent external perturbation $V_{\rm ext}(t)$ is present, the density deviation obeys the so-called time-dependent Hartree(-Fock) equation.
In the frequency representation, it is expressed as
\begin{equation}\label{eq:linear}
    \omega\delta\rho(\omega)=[h_0,\delta\rho(\omega)]+[\delta h(\omega)+V_{\rm ext}(\omega),\rho_0],
\end{equation}
as a linear response to the weak perturbation.

In practical calculations, it is convenient to adopt the single-particle orbitals to represent the density matrix, $\rho(t)=\sum_{i=1}^A \lr\psi_i(t)\rc\lc\psi_i(t)\rl$.
The density deviation in the frequency representation can be expressed as
\begin{equation}
  \delta\rho(\omega) = \sum_{i=1}^A \{\lr X_i(\omega)\rc\lc\phi_i\rl+\lr\phi_i\rc\lc Y_i(\omega)\rl\},
\end{equation}
with the so-called forward $X(\omega)$ and backward $Y(\omega)$ amplitudes.
It is slightly tricky that one must take the ket $\lr X_i(\omega)\rc$ and bra $\lc Y_i(\omega)\rl$ states independent since $\delta\rho(\omega)$ is not Hermitian.

In conventional RPA calculations, the $X(\omega)$ and $Y(\omega)$ amplitudes are expanded on the basis of particle states,
$\lr X_i(\omega)\rc=\sum_{m>A}\lr\phi_m\rc X_{mi}(\omega)$ and $\lr Y_i(\omega)\rc=\sum_{m>A}\lr\phi_m\rc Y^*_{mi}(\omega)$.
Then, one can derive the well-known RPA equation in the matrix form,
\begin{align}\label{eq:RPA}
    &\Lb\lb\begin{array}{cc}
      \mathcal{A}_{mi,nj} & \mathcal{B}_{mi,nj} \\
      \mathcal{B}^*_{mi,nj} & \mathcal{A}^*_{mi,nj}
    \end{array}\rb
    -\omega
    \lb\begin{array}{cc}
      1 & 0 \\
      0 & -1
    \end{array}\rb\Rb
    \lb\begin{array}{c}
      X_{nj} \\
      Y_{nj}
    \end{array}\rb\nonumber\\
    &=-\lb\begin{array}{c}
      f_{mi} \\
      g_{mi}
    \end{array}\rb.
\end{align}
The RPA matrices $\mathcal{A}$ and $\mathcal{B}$ read
\begin{subequations}\label{eq:AB}
\begin{align}
    \mathcal{A}_{mi,nj}
        &=(\epsilon_m-\epsilon_i)\delta_{mn}\delta_{ij}+\mean{\phi_m}{\left.\frac{\partial h}{\partial\rho_{nj}}\right|_{\rho=\rho_0}}{\phi_i}\nonumber\\
        &=(\epsilon_m-\epsilon_i)\delta_{mn}\delta_{ij}+\mean{\phi_m\phi_j}{V_{ph}}{\phi_n\phi_i},\\
    \mathcal{B}_{mi,nj}
        &=\mean{\phi_m}{\left.\frac{\partial h}{\partial\rho_{jn}}\right|_{\rho=\rho_0}}{\phi_i}
         =\mean{\phi_m\phi_n}{V_{ph}}{\phi_j\phi_i}.
\end{align}
\end{subequations}

$\bigstar$ Rearrangement terms: the \textit{ph} residual interactions $V_{ph}$ in the fully self-consistent calculations are strictly derived from the second derivative of the energy functional.
The density-dependent coupling strengths $\alpha$ introduces the rearrangement terms in $V_{ph}$ with $\partial\alpha/\partial\rho_b$ or $\partial^2\alpha/\partial\rho_b^2$ \cite{Niksic2002}.
They are calculated separately in the conventional RPA calculations.

$\bigstar$ Dirac sea: the relativistic RPA is equivalent to the time-dependent RMF theory in the small amplitude limit, only when the particle states $m,n$ include not only the states above the Fermi surface but
also the states in the Dirac sea \cite{Ring2001}.
It is due to the no-sea approximation used in the ground-state calculations.

\subsection{Iterative finite-amplitude method}\label{sec:iFAM}

In Ref.~\cite{Nakatsukasa2007}, the FAM was proposed as a simpler and more efficient approach to the solutions of the linear response equation~(\ref{eq:linear}).
This method does not require explicit evaluation of the residual interactions $\delta h/\delta\rho$ (\ref{eq:AB}).
Instead, by multiplying with the ket $\lr\phi_i\rc$ and bra $\lc\phi_i\rl$ of only hole states on both sides of Eq.~(\ref{eq:linear}), respectively, one has
\begin{subequations}\label{eq:FAM}
\begin{align}
    \omega\lr X_i(\omega)\rc
        &= (h_0-\epsilon_i)\lr X_i(\omega)\rc+\hat Q(V_{\rm ext}(\omega)+\delta h(\omega))\lr\phi_i\rc,\\
    \omega^*\lr Y_i(\omega)\rc
        &= -(h_0-\epsilon_i)\lr Y_i(\omega)\rc-\hat Q(V^\dag_{\rm ext}(\omega)+\delta h^\dag(\omega))\lr\phi_i\rc.
\end{align}
\end{subequations}

The $\delta h(\omega)$ and $\delta h^\dag(\omega)$ are calculated by using the finite difference with a sufficiently small number $\eta$:
\begin{equation}\label{eq:dh}
  \delta h(\omega)=\ff\eta(h[\lc\psi'\rl,\lr\psi\rc]-h[\lc\phi\rl,\lr\phi\rc])
\end{equation}
with $\lc\psi'_i\rl=\lc\phi_i\rl+\eta\lc Y_i(\omega)\rl, \lr\psi_i\rc=\lr\phi_i\rc+\eta\lr X_i(\omega)\rc$, and
\begin{equation}\label{eq:dh+}
  \delta h^\dag(\omega)=\ff\eta(h[\lc\psi'\rl,\lr\psi\rc]-h[\lc\phi\rl,\lr\phi\rc])
\end{equation}
with
$\lc\psi'_i\rl=\lc\phi_i\rl+\eta\lc X_i(\omega)\rl, \lr\psi_i\rc=\lr\phi_i\rc+\eta\lr Y_i(\omega)\rc$.

In the coordinate space, for instance, the corresponding radial FAM equations for the monopole responses read
\begin{subequations}\label{eq:iFAM}
\begin{align}
    &\hat Q\ls(h_0(r)-\epsilon_i-\omega)X_i(r,\omega) +\delta h(r,\omega)\phi_i(r)\rs\no
    =&  -\hat Q V_{\rm ext}(r,\omega)\phi_i(r),\\
    &\hat Q\ls(h_0(r)-\epsilon_i+\omega^*)Y_i(r,\omega) +\delta h^\dag(r,\omega)\phi_i(r)\rs^*\no
    =& -\hat Q \ls V^\dag_{\rm ext}(r,\omega)\phi_i(r)\rs^*.
\end{align}
\end{subequations}
This FAM equation is a standard linear algebraic equation of the form, $\mathcal{A}\vec x=\vec b$, which can be solved within the iterative scheme.
In such a way, we do not need to construct the matrix elements of $\mathcal{A}$ explicitly, but only to evaluate $\mathcal{A}\vec x$ for a given vector $\vec x$.
This is called the iterative finite-amplitude method (i-FAM).

The practical procedure for evaluating $\delta h(r)$ and $\delta h^\dag(r)$ is following:
with a given set of $\{X_i(r)\}$ and $\{Y_i(r)\}$, one sequentially calculates
\begin{itemize}
  \item[---] the nucleon densities and currents in Eq.~(\ref{eq:dens});
  \item[---] the new coupling strengths in Eq.~(\ref{eq:alpha});
  \item[---] the Coulomb fields in Eq.~(\ref{eq:Coulomb});
  \item[---] the rearrangement self-energy in Eq.~(\ref{eq:SigmaR});
  \item[---] the scalar and vector potentials in Eq.~(\ref{eq:Sigma});
  \item[---] the one-body Hamiltonian $h(r)$ in Eq.~(\ref{eq:rDirac}).
\end{itemize}

$\bigstar$ Since now the $X(r)$ and $Y(r)$ amplitudes are independent due to the non-Hermitian nature of $\delta h(r)$ and $\delta h^\dag(r)$, it is clear that the nucleon currents are no longer vanishing.
This is the reason why the time-odd terms must be kept from the beginning.

$\bigstar$ Note that here the $X$ and $Y$ amplitudes are expanded on the mesh points $\{r_k\}$ in the coordinate space.
In such a way, even though the effects of the Dirac sea cannot be identified or isolated, these effects are properly taken into
account, because the coordinate space $\sum_{\mathbf{r}}\lr\mathbf{r}\rc\lc\mathbf{r}\rl-\sum_j\lr\phi_j\rc\lc\phi_j\rl$ provides a complete set of basis for particle states.

$\bigstar$ In order to include both the normal and rearrangement terms in $V_{ph}$, one simply needs to re-calculate the coupling strengths $\alpha$ and their derivatives $\partial\alpha/\partial\rho_b$ by using Eq.~(\ref{eq:alpha}) for each given set of $\{X_i(r)\}$ and $\{Y_i(r)\}$.
If one skips this step, this means the normal terms in $V_{ph}$ remain, but all of the rearrangement terms are neglected.

\begin{figure}[t]
\includegraphics[width=8cm]{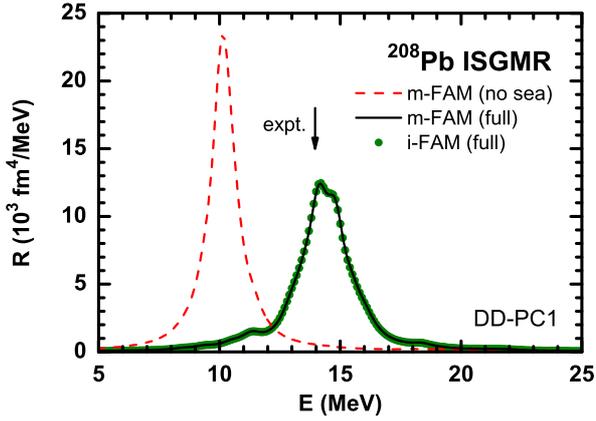}
\caption{(Color online) ISGMR in $^{208}$Pb calculated by i-FAM and m-FAM.
The i-FAM results are shown with the dotted symbols, while the m-FAM results calculated with and without the Dirac sea are shown with the solid and dashed lines, respectively.
The experimental centroid energy \cite{Youngblood2004} is denoted by the arrow.
Taken from Ref.~\cite{Liang2013}.
\label{Fig1}}
\end{figure}

\begin{figure}[t]
\includegraphics[width=8cm]{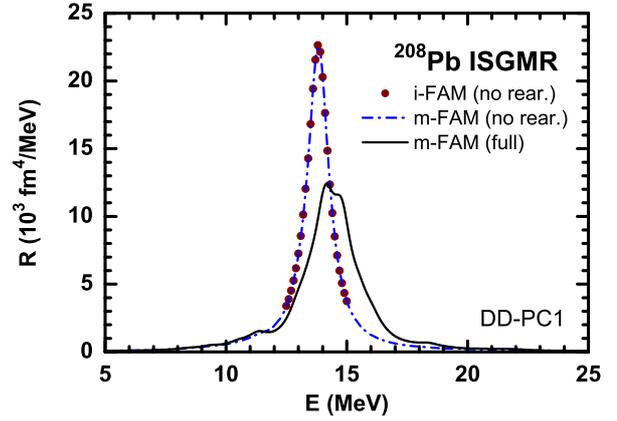}
\caption{(Color online) ISGMR in $^{208}$Pb calculated by i-FAM and m-FAM.
The i-FAM results without the rearrangement terms are shown with the dotted symbols, while the m-FAM results calculated with and without the rearrangement terms are shown with the solid and dash-dotted lines, respectively.
Taken from Ref.~\cite{Liang2013}.
\label{Fig2}}
\end{figure}

\subsection{Matrix finite-amplitude method}\label{sec:mFAM}

Another usage of FAM is the matrix finite-amplitude method (m-FAM) \cite{Avogadro2013}.
In this method, the RPA matrices $\mathcal{A}$ and $\mathcal{B}$ are explicitly constructed, but the tedious calculations concerning $V_{ph}$ can be avoided.

The kernels $\partial h/\partial\rho$ in Eq.~(\ref{eq:AB}) are directly calculated with the finite difference \begin{equation}
  \left.\frac{\partial h}{\partial\rho_{nj}}\right|_{\rho=\rho_0} =
  \ff\eta(h[\lc\psi'\rl,\lr\psi\rc]-h[\lc\phi\rl,\lr\phi\rc]),
\end{equation}
by keeping all $\lc\psi'_i\rl=\lc\phi_i\rl$ and $\lr\psi_i\rc=\lr\phi_i\rc$ unchanged, but slightly mixing specific orbitals $j$ with $n$ as $\lr\psi_j\rc=\lr\phi_j\rc+\eta\lr \phi_n\rc$.
In the same way,
\begin{equation}
  \left.\frac{\partial h}{\partial\rho_{jn}}\right|_{\rho=\rho_0} =
  \ff\eta(h[\lc\psi'\rl,\lr\psi\rc]-h[\lc\phi\rl,\lr\phi\rc]),
\end{equation}
with a small mixing of the specific orbitals $j$ and $n$ as $\lc\psi'_j\rl=\lc\phi_j\rl+\eta\lc \phi_n\rl$.

$\bigstar$ To include the effects of the Dirac sea, states $n$ run over the unoccupied states in both Fermi and Dirac sea.
To include the effects of the rearrangement terms, one follows the same procedure as that in i-FAM shown above.

\section{Illustrative calculations}

As illustrative calculations, we show the isoscalar giant monopole resonances (ISGMR) in $^{208}$Pb, where the density-dependent point-coupling RMF parametrization DD-PC1 \cite{Niksic2008} is used and the spherical symmetry is assumed.
The effects of the Dirac sea and the rearrangement terms can be examined by switching on or off the corresponding \textit{ph} residual interactions.

First of all, the transition strengths of ISGMR in $^{208}$Pb calculated in the m-FAM scheme with and without the Dirac sea are compared in Fig.~\ref{Fig1}.
It is found that the Dirac sea shows profound effects on the centroid energy and the experimental data \cite{Youngblood2004} is reproduced only when the Dirac sea is taken into account.

As mentioned before, the effects of the Dirac sea cannot be isolated in the coordinate-space representation.
However, it can be clearly seen in Fig.~\ref{Fig1} that the i-FAM results are exactly on top of the m-FAM results that include the Dirac sea.
This confirms that the coordinate space generates another complete set of basis for particle states and these two FAM schemes are equivalent.

For the rearrangement terms, it is tedious to calculate them in the conventional RPA calculations, in contrast, in FAM these terms can be simply taken into account by re-calculating the coupling strengths $\alpha$ and their derivatives $\partial\alpha/\partial\rho_b$ with Eq.~(\ref{eq:alpha}).
The numerical cost of such a step is totally negligible.

In Fig.~\ref{Fig2}, the transition strengths of ISGMR in $^{208}$Pb calculated by m-FAM with and without the rearrangement terms are shown, together with the i-FAM results calculated without the rearrangement terms.
The equivalency of these two FAM schemes is demonstrated once more.
Quantitatively, it is found that the rearrangement effects on the centroid energies is also substantial.


\section{Summary}

Work is now in progress for establishing the self-consistent relativistic RPA by using the FAM.
It is shown that, in the present schemes, the effects of the Dirac sea can be automatically taken into account in the coordinate-space representation.
The rearrangement terms due to the density-dependent couplings can be also implicitly calculated without extra computational costs.

\begin{ack}
This work was partly supported by a Grant-in-Aid for JSPS Fellows under Grant No. 24-02201,
the JSPS KAKENHI under Grants No. 24105006, No. 25287065, and No. 25287066,
the Major State 973 Program 2013CB834400,
the National Natural Science Foundation of China under Grants No. 10975008, No. 11105006, No. 11175002, and No. 11205004,
and the Research Fund for the Doctoral Program of Higher Education under Grant No. 20110001110087.
\end{ack}



\end{document}